\RequirePackage{fix-cm}

\documentclass[twocolumn,epjc3]{svjour3}
\usepackage[usenames]{color}
\usepackage{color}

\RequirePackage[T1]{fontenc}

\RequirePackage{graphicx} \RequirePackage{subfigure}
\RequirePackage{rotating} \RequirePackage{pdflscape}
\RequirePackage{color} \RequirePackage{lineno}

\journalname{Eur. Phys. J. C}

\newcommand{\isomo}{$^{100}$Mo}
\newcommand{\isoru}{$^{100}$Ru}
\newcommand{\zeronu}{$0\nu\beta\beta$}
\newcommand{\twonu}{$2\nu\beta\beta$}
\newcommand{\crystal}{Li$_2$$^{100}$MoO$_4$}
\newcommand{\crystalplain}{Li$_2$MoO$_4$}
\newcommand{\etal}{\emph{et al.}}
\newcommand{\THL}{\ensuremath{T_{1/2}^{2\nu}}}

\begin{document}

\title{Precise measurement of $2\nu\beta\beta$ decay of $^{100}$Mo with the CUPID-Mo detection technology}

\author{
 E.~Armengaud\thanksref{CEA-IRFU}\and
 C.~Augier\thanksref{IPNL}\and
 A.S.~Barabash\thanksref{ITEP}\and
 F.~Bellini\thanksref{Sapienza,INFN-Roma}\and
G.~Benato\thanksref{UCB,LBNL} \and
 A.~Beno\^{\i}t\thanksref{Neel}\and
 M.~Beretta\thanksref{Milano,INFN-Milano}\and
 L.~Berg\'e\thanksref{CSNSM}\and
 J.~Billard\thanksref{IPNL}\and
 Yu.A.~Borovlev\thanksref{NIIC}\and
 Ch.~Bourgeois\thanksref{LAL}\and
 M.~Briere\thanksref{LAL}\and
 V.~Brudanin\thanksref{JINR}\and
 P.~Camus\thanksref{Neel}\and
 L.~Cardani\thanksref{INFN-Roma}\and
 N.~Casali\thanksref{INFN-Roma}\and
 A.~Cazes\thanksref{IPNL}\and
 M.~Chapellier\thanksref{CSNSM}\and
 F.~Charlieux\thanksref{IPNL}\and
 M.~de~Combarieu\thanksref{CEA-IRAMIS}\and
 I.~Dafinei\thanksref{INFN-Roma}\and
 F.A.~Danevich\thanksref{KINR}\and
 M.~De~Jesus\thanksref{IPNL}\and
 L.~Dumoulin\thanksref{CSNSM}\and
 K.~Eitel\thanksref{KIT-IK}\and
 E.~Elkhoury\thanksref{IPNL}\and
 F.~Ferri\thanksref{CEA-IRFU}\and
 B.K.~Fujikawa\thanksref{LBNL}\and
 J.~Gascon\thanksref{IPNL}\and
 L.~Gironi\thanksref{Milano,INFN-Milano}\and
 A.~Giuliani\thanksref{e1,CSNSM}\and
 V.D.~Grigorieva\thanksref{NIIC}\and
 M.~Gros\thanksref{CEA-IRFU}\and
 E.~Guerard\thanksref{LAL}\and
 D.L.~Helis\thanksref{CEA-IRFU}\and
 H.Z.~Huang\thanksref{Fudan}\and
 R.~Huang\thanksref{UCB}\and
 J.~Johnston\thanksref{MIT}\and
 A.~Juillard\thanksref{IPNL}\and
 H.~Khalife\thanksref{CSNSM}\and
 M.~Kleifges\thanksref{KIT-IPE}\and
 V.V.~Kobychev\thanksref{KINR}\and
 Yu.G.~Kolomensky\thanksref{UCB,LBNL}\and
 S.I.~Konovalov\thanksref{ITEP}\and
 A.~Leder\thanksref{MIT}\and
 J.~Kotila\thanksref{FIER}\and
 P.~Loaiza\thanksref{LAL}\and
 L.~Ma\thanksref{Fudan}\and
 E.P.~Makarov\thanksref{NIIC}\and
 P.~de~Marcillac\thanksref{CSNSM}\and
L.~Marini\thanksref{UCB}\and
 S.~Marnieros\thanksref{CSNSM}\and
 D.~Misiak\thanksref{IPNL}\and
 X-F.~Navick\thanksref{CEA-IRFU}\and
 C.~Nones\thanksref{CEA-IRFU}\and
 V.~Novati\thanksref{CSNSM}\and
 E.~Olivieri\thanksref{CSNSM}\and
 J.L.~Ouellet\thanksref{MIT}\and
 L.~Pagnanini\thanksref{Milano,INFN-Milano}\and
 P.~Pari\thanksref{CEA-IRAMIS}\and
 L.~Pattavina\thanksref{LNGS,TUM}\and
 B.~Paul\thanksref{CEA-IRFU}\and
 M.~Pavan\thanksref{Milano,INFN-Milano}\and
 H.~Peng\thanksref{USTC}\and
 G.~Pessina\thanksref{INFN-Milano}\and
 S.~Pirro\thanksref{LNGS}\and
 D.V.~Poda\thanksref{CSNSM,KINR}\and
 O.G.~Polischuk\thanksref{KINR}\and
 E.~Previtali\thanksref{Milano,INFN-Milano}\and
 Th.~Redon\thanksref{CSNSM}\and
 S.~Rozov\thanksref{JINR}\and
 C.~Rusconi\thanksref{USC}\and
 V.~Sanglard\thanksref{IPNL}\and
 K.~Sch\"affner\thanksref{LNGS}\and
 B.~Schmidt\thanksref{LBNL}\and
 Y.~Shen\thanksref{Fudan}\and
 V.N.~Shlegel\thanksref{NIIC}\and
 B.~Siebenborn\thanksref{KIT-IK}\and
 V.~Singh\thanksref{UCB}\and
 C.~Tomei\thanksref{INFN-Roma}\and
 V.I.~Tretyak\thanksref{KINR}\and
 V.I.~Umatov\thanksref{ITEP}\and
 L.~Vagneron\thanksref{IPNL}\and
 M.~Vel\'azquez\thanksref{UGA}\and
 M.~Weber\thanksref{KIT-IPE}\and
 B.~Welliver\thanksref{LBNL}\and
 L.~Winslow\thanksref{MIT}\and
 M.~Xue\thanksref{USTC}\and
 E.~Yakushev\thanksref{JINR}\and
 A.S.~Zolotarova\thanksref{CEA-IRFU,CSNSM}
 }

\thankstext{e1}{e-mail: andrea.giuliani@csnsm.in2p3.fr}

\institute{
 IRFU, CEA, Universit\'{e} Paris-Saclay, F-91191 Gif-sur-Yvette, France  \label{CEA-IRFU} \and
 Univ Lyon, Universit\'{e} Lyon 1, CNRS/IN2P3, IP2I-Lyon, F-69622, Villeurbanne, France  \label{IPNL} \and
 National Research Centre Kurchatov Institute, Institute of Theoretical and Experimental Physics, 117218 Moscow, Russia \label{ITEP} \and
 Dipartimento di Fisica, Sapienza Universit\`a di Roma, P.le Aldo Moro 2, I-00185, Rome, Italy \label{Sapienza} \and
 INFN, Sezione di Roma, P.le Aldo Moro 2, I-00185, Rome, Italy \label{INFN-Roma} \and
 Department of Physics, University of California, Berkeley, California 94720, USA \label{UCB} \and
 Nuclear Science Division, Lawrence Berkeley National Laboratory, Berkeley, CA 94720, USA \label{LBNL} \and
 CNRS-N\'eel, 38042 Grenoble Cedex 9, France \label{Neel} \and
 CSNSM, Univ. Paris-Sud, CNRS/IN2P3, Universit\'e Paris-Saclay, 91405 Orsay, France \label{CSNSM} \and
 Nikolaev Institute of Inorganic Chemistry, 630090 Novosibirsk, Russia \label{NIIC} \and
 LAL, Univ. Paris-Sud, CNRS/IN2P3, Universit\'e Paris-Saclay, 91898 Orsay, France \label{LAL} \and
 Laboratory of Nuclear Problems, JINR, 141980 Dubna, Moscow region, Russia \label{JINR} \and
 IRAMIS, CEA, Universit\'{e} Paris-Saclay, F-91191 Gif-sur-Yvette, France \label{CEA-IRAMIS} \and
 Institute for Nuclear Research, 03028 Kyiv, Ukraine \label{KINR} \and
 Karlsruhe Institute of Technology, Institut f\"{u}r Kernphysik, 76021 Karlsruhe, Germany \label{KIT-IK} \and
 Dipartimento di Fisica, Universit\`{a} di Milano - Bicocca, I-20126 Milano, Italy \label{Milano} \and
 INFN, Sezione di Milano - Bicocca, I-20126 Milano, Italy \label{INFN-Milano} \and
 Key Laboratory of Nuclear Physics and Ion-beam Application (MOE), Fudan University, Shanghai 200433, PR China \label{Fudan} \and
 Massachusetts Institute of Technology, Cambridge, MA 02139, USA \label{MIT} \and
 Karlsruhe Institute of Technology, Institut f\"{u}r Prozessdatenverarbeitung und Elektronik, 76021 Karlsruhe, Germany \label{KIT-IPE} \and
 Finnish Institute for Educational Research, University of Jyv\"{a}skyl\"{a}, P.O. Box 35, Jyv\"{a}skyl\"{a} FI-40014, Finland \label{FIER} \and
 INFN, Laboratori Nazionali del Gran Sasso, I-67100 Assergi (AQ), Italy \label{LNGS} \and
 Physik Department, Technische Universit\"at M\"unchen, Garching D-85748, Germany \label{TUM} \and
 Department of Physics and Astronomy, University of South Carolina, SC 29208, Columbia, USA \label{USC} \and
 Department of Modern Physics, University of Science and Technology of China, Hefei 230027, PR China \label{USTC} \and
 Universit\'e Grenoble Alpes, CNRS, Grenoble INP, SIMAP, 38402 Saint Martin d'H\'eres, France \label{UGA}
 }
\date{Received: date  / Accepted: date}
\maketitle

\begin{abstract}
We report the measurement of the two-neu\-trino double-beta
($2\nu\beta\beta$) decay of
$^{100}$Mo to the ground state of $^{100}$Ru using lithium molybdate
(\crystal) scintillating bolometers. The detectors were developed for the
CUPID-Mo program and operated at the EDEL\-WEISS-III low background
facility in the Modane underground laboratory.  From a total exposure
of $42.235$ kg$\times$d, the half-life of $^{100}$Mo is determined to be
$T_{1/2}^{2\nu}=[7.12^{+0.18}_{-0.14}\,\mathrm{(stat.)}\pm0.10\,\mathrm{(syst.)}]\times10^{18}$~years. This is the most accurate determination of the
 $2\nu\beta\beta$ half-life of $^{100}$Mo to date. We also confirm, with the
statistical significance of $>3\sigma$, that the single-state
dominance model of the $2\nu\beta\beta$ decay of $^{100}$Mo is favored over the
high-state dominance model.
\end{abstract}


\section{Introduction}
\label{sec:Intro}

Two-neutrino double-beta (\twonu) decay and the related process of the
two-neutrino double-electron capture (2$\nu$ECEC) are allowed
second-order processes in
the Standard Model theory of electroweak interactions and are the
rarest nuclear processes ever observed
\cite{Tretyak:2002,Saakyan:2013,Barabash:2019}. Precise
measurements of these processes are critical to understanding
the nuclear physics governing \twonu\ and to benchmarking the
calculations of the beyond the Standard
Model process, zero-neutrino double-beta (\zeronu) decay. The
observation of the latter process would establish the Majorana nature
of the neutrino and is therefore the subject of a global experimental
effort.

Double-beta decay is observable in nuclei where the single beta decay
is forbidden or highly suppressed. Of the candidate isotopes,
\isomo\ is characterized by one
of the largest decay energies ($Q_{\beta\beta} = 3034.36(17)$ \,keV)
\cite{Wang:2017} and the shortest \twonu\ half-life
\cite{Barabash:2019}.  Table~\ref{tab:experiments}
summarizes the measurements of the \isomo\, \twonu\ half-life to
date. Most experiments have used \isomo\ foils coupled with traditional tracking
and calorimetry techniques. NEMO-3 presents the most precise
measurement to date at
$T_{1/2}^{2\nu}=[6.81\pm0.01(\mathrm{stat.})^{+0.38}_{-0.40}(\mathrm{syst.})]$\,yr
\cite{Arnold:2019}. The separate foil and detector design limits the
scalability of the experiment to large isotope masses; most of the leading
\zeronu~experiments are moving towards the combined detector and
isotope design. The most precise previous measurement using the
``source = detector'' approach is
$\THL=[7.15\pm0.37(\mathrm{stat.})\pm0.66(\mathrm{syst.})]\times10^{18}$~yr
\cite{Cardani:2014} using zinc molybdate (ZnMoO$_4$) crystals operated as
scintillating bolometers.

The bolometric technique is now competitive with
the foil-based detectors, and offers distinct advantages. In this
work, \crystal~crystals operated as
scintillating bolometers and developed as part of the CUPID-Mo program
\cite{Armengaud:2017b,Poda:2017,Zolotarova:2018,Poda:2018,Armengaud:2019} are used to precisely
measure the \isomo~half-life. CUPID-Mo is a demonstrator experiment
for CUPID~\cite{Wang:2015,Armstrong:2019}, a proposed
next-generation bolometric search for \zeronu\ in \isomo\
at the Laboratori
Nazionali del Gran Sasso (LNGS). CUPID will
use the infrastructure built for the CUORE \zeronu\
experiment~\cite{Alduino:2018},
currently in operation at LNGS.

\begin{table*}[ht]
\caption{Measurements of $2\nu\beta\beta$ decay of $^{100}$Mo to date.}
\begin{center}
\begin{tabular}{l|l|l}
 \hline
 Description                                                    & $\THL (\times 10^{18}$ yr)                                 & Year, Reference\\
 \hline
 ELEGANT V: $^{100}$Mo and $^{\mathrm{nat}}$Mo foils,           & ~                                                             & ~ \\
 drift chambers, plastic scintillators                          & $11.5^{+3.0}_{-2.0}$                                          & 1991 \cite{Ejiri:1991}  \\
 \hline
 NEMO-2: $^{100}$Mo foil, track reconstruction by               & $9.5\pm0.4(\mathrm{stat.})\pm0.9(\mathrm{syst.})$             & 1995 \cite{Dassie:1995} \\
 Geiger cells, plastic scintillators                            & $7.51\pm0.28(\mathrm{stat.})^{+0.53}_{-0.31}(\mathrm{syst.})$ & 1997 \cite{Vareille:1997} \\
 \hline
 $^{100}$Mo foil, segmented Si(Li) detector                     & $7.6^{+2.2}_{-1.4}$                                           & 1997 \cite{Alston:1997}  \\
 \hline
 Hoover Dam: $^{100}$Mo foil, time-projection                   & ~ & ~ \\
 chamber                                                        & $6.82^{+0.38}_{-0.53}(\mathrm{stat.})\pm0.68(\mathrm{syst.})$ & 1997 \cite{DeSilva:1997} \\
 \hline
 DBA: $^{100}$Mo foil, liquid argon ionization                  & ~             & ~ \\
 chamber                                                        & $7.2\pm0.9(\mathrm{stat.})\pm1.8(\mathrm{syst.})$             & 2001 \cite{Ashitkov:2001} \\
 \hline
 Geochemical, isotope dilution mass spectrometry                & ~                                                             & ~ \\
 of old molybdenites                                            & $2.1\pm0.3$                                                   & 2004 \cite{Hidaka:2004} \\
 \hline
 NEMO-3: $^{100}$Mo foil, track reconstruction by               & ~                                                             & ~  \\
 Geiger cells, plastic scintillators                            & $7.11\pm0.02(\mathrm{stat.})\pm0.54(\mathrm{syst.})$          & 2005 \cite{Arnold:2005}  \\
 \hline
 Low temperature ZnMoO$_4$ bolometers                           & $7.15\pm0.37(\mathrm{stat.})\pm0.66(\mathrm{syst.})$          & 2014 \cite{Cardani:2014}   \\
 \hline
 NEMO-3: $^{100}$Mo foil, track reconstruction by               & ~                                                             & ~  \\
 Geiger cells, plastic scintillators                            & $6.81\pm0.01(\mathrm{stat.})^{+0.38}_{-0.40}(\mathrm{syst.})$ & 2019 \cite{Arnold:2019}  \\
 \hline
\end{tabular}
\end{center}
\label{tab:experiments}
\end{table*}

\section{Experiment}
\label{sec:exp}

This measurement uses four lithium molybdate crystals enriched in
\isomo~\, (\crystal)
instrumented as low temperature scintillating bolometers.
The crystals were
produced as part of the LUMINEU project~\cite{LUMINEU}. They were
grown using the
low-thermal-gradient Czochralski technique starting from the highly
purified enriched molybdenum oxide and lithium carbonate
\cite{Grigorieva:2017}. The R\&D of large volume \crystalplain~based
scintillating bolometers is described in
\cite{Poda:2017,Armengaud:2017b,Bekker:2016}. The \isomo~enrichment of
the molybdenum oxide precursor varied slightly between the different
crystal productions. The final enrichment fraction in the
  grown crystals and its uncertainty was estimated taking into account
  the uncertainties
in the original precursor enrichment ($\pm$0.05\%) and the effect of
the crystal growth process. Table~\ref{tab:crystals} summarizes the
crystal dimensions, masses, isotopic abundance of
  $^{100}$Mo in the crystals, and number of \isomo~nuclei.

\nopagebreak
\begin{table*}[ht]
\caption{\crystal\ crystal scintillators used in the experiment.}
\begin{center}
\begin{tabular}{l|l|l|l|r|r}
 \hline
 Crystal & Crystal mass (g),             & \isomo\ isotopic  & Number of             & \multicolumn{2}{c}{Live time (h)}  \\
 \cline{5-6}
 number  & size (mm)                     & abundance (\%)  & \isomo~nuclei        & setup 1  & setup 2 \\
 \hline
 1 & 185.86(1), $\oslash 43.6\times40.0$    & 96.93(7)         & $6.105(9)\times10^{23}$  & 1331.03   & 1000.58 \\
 2 & 203.72(1), $\oslash 43.6\times44.2$    & 96.93(7)         & $6.692(10)\times10^{23}$  & ~         & 997.64 \\
 3 & 212.61(1), $\oslash 43.9\times45.6$    & 96.89(12)        & $6.981(16)\times10^{23}$  & ~         & 1037.92 \\
 4 & 206.68(1), $\oslash 43.9\times44.5$    & 96.89(12)        & $6.786(15)\times10^{23}$  & ~         & 756.59 \\
 \hline
\end{tabular}
\end{center}
\label{tab:crystals}
\end{table*}

\normalsize

Each \crystal~crystal is instrumented with a neutron transmutation
doped (NTD) germanium temperature sensor \cite{Haller:1994} and a
heavily-doped silicon heater. The latter is used to stabilize the
  thermal response of the detector \cite{Andreotti:2012}. The two devices are
  glued to the crystal surface and then the
crystals are installed in a copper holder and secured by PTFE support
clamps. A light detector constructed from a Germanium disc
$\oslash44\times0.17$ mm instrumented with an NTD sensor is installed
above each crystal to detect the scintillation signal from the
crystal. The simultaneous detection of heat and light signals provides
a powerful discrimination between $\gamma$($\beta$) and $\alpha$
events~\cite{Poda:2017c}. This discrimination is key
in the analysis that follows for both the estimation and reduction of
backgrounds.

\nopagebreak
\begin{figure*}[htbp]
\centering
\includegraphics[width=1.0\textwidth]{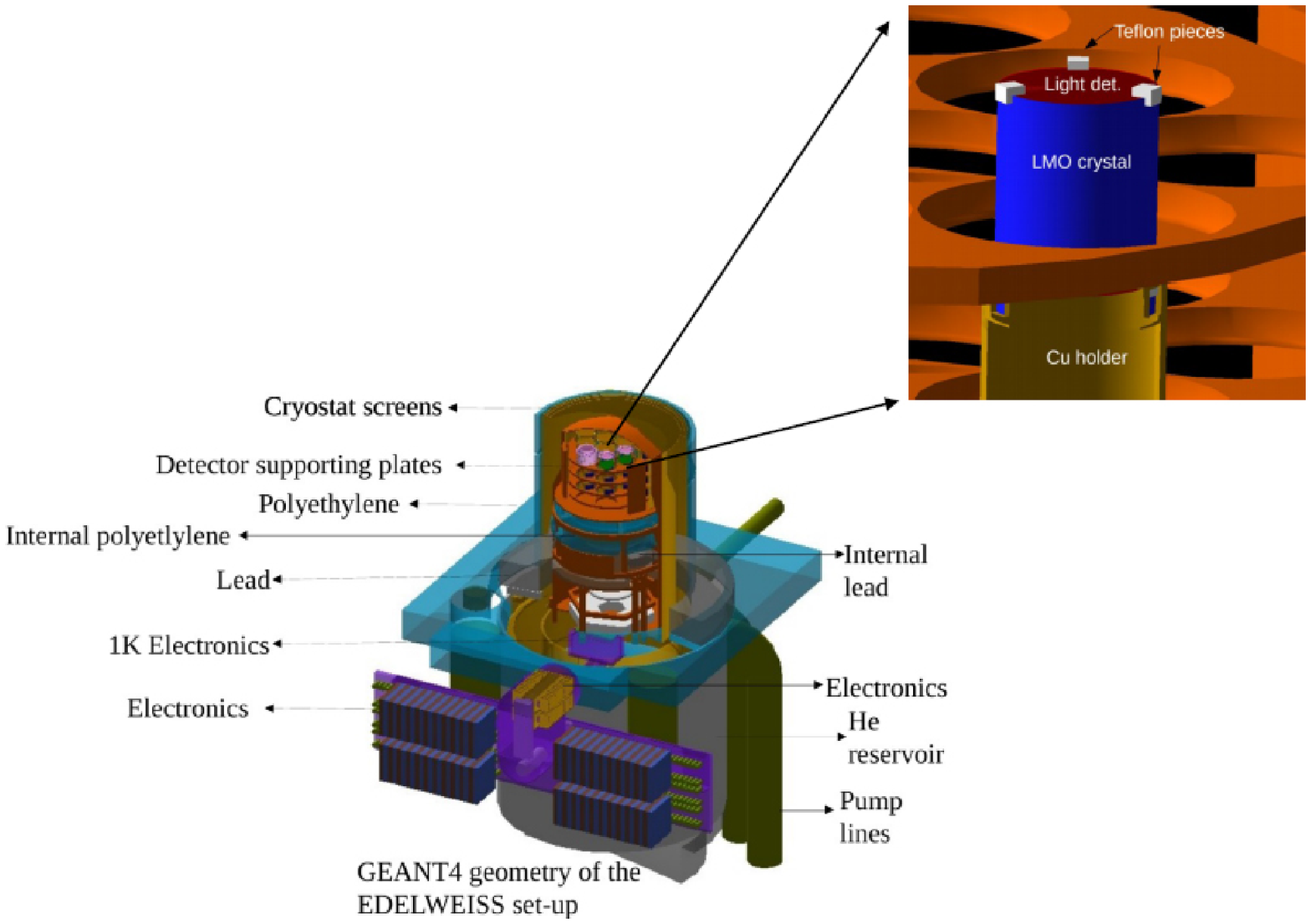}
\caption{EDELWEISS-III cryostat modeled in {\sc Geant4} as
configured for the experiment. (Zoom) The individual detectors are
constructed from a crystal supported in a copper holder that also
holds the light detector which is constructed with a Germanium
wafer. The details of the copper assembly surrounding
the crystals are not shown.} \label{fig:Exp}
\end{figure*}

\begin{figure}[htbp]
\centering
\includegraphics[width=0.48\textwidth]{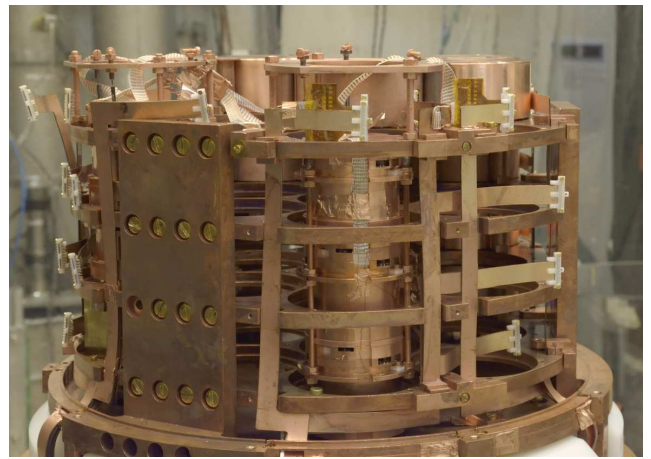}
\caption{EDELWEISS
cryogenic facility with partially installed
detector modules.} \label{fig:pic}
\end{figure}

The experiment operated in the low-background cryostat of the
EDELWEISS-III dark-matter experiment \cite{Armengaud:2017a}, see
Figs.~\ref{fig:Exp}-\ref{fig:pic}. The cryostat is located in the Modane underground
laboratory (France) at the depth of 4800 m of water equivalent. The
central volume of the EDEL\-WEISS-III cryostat is shielded by 20\,cm
of Pb, the innermost 2\,cm is Roman Pb to reduce the $^{210}$Pb
background contribution. The experiment was realized in two steps: a
single crystal configuration, ``setup 1'', and a four-crystal
configuration, ``setup 2''. The detector modules and materials in the
two setups were slightly different, producing a somewhat different
background composition. EDELWEISS germanium detectors were run
concurrently with this measurement.

In addition to the differences in geometry and materials
between the two detector configurations, there was a change in the data
acquisition during setup 2. Setup 1 and $\approx$22\% of setup 2 were
triggered online, while the remainder of setup 2 was acquired in the
streaming data acquisition mode and then triggered offline.
The data acquired during instabilities of the cryogenic system were
not used for the analysis. If the temperature of the detector holder
plate showed variations larger than $\pm0.1$~mK from a chosen value
($20.0$~mK and $19.2$~mK for setup 1, and $17.0$~mK for setup 2), the
data were discarded. Similarly, we discard periods of large
non-thermal variations in the detector baselines.
As a result, $\sim$7\% and $\sim$12\% of physics data were not
  considered in the present analysis for setup 1 and 2, respectively.
Table~\ref{tab:crystals} summarizes the live-time for each
configuration. The uncertainty in the live-time calculation is
estimated from the loss of the periodically injected heater
signals. This uncertainty for the online-triggered data is $0.23\%$ and the
uncertainty in the stream mode is $0.22\%$, leading to the exposure-weighted
average of $0.22\%$.

The energy scale and energy resolution of the detectors are calibrated
using $^{40}$K, $^{133}$Ba, and $^{232}$Th gamma
sources~\cite{Poda:2017,Armengaud:2017b}. The energy
resolution is measured at $356.0$~keV ($^{133}$Ba), $1460.8$~keV
($^{40}$K) and $2614.5$~keV ($^{208}$Tl) resulting in $\sim3$ keV,
$\sim5$ keV and $\sim6$ keV full width at half maximum
(FWHM), respectively.
The energy scale is stable to within $\pm0.12$\%
as determined from the variation observed in the
periodic $^{133}$Ba calibrations and the physics data ($^{210}$Po
$\alpha$ events originating in the crystal bulk).
After applying the energy calibration, we observe a modest residual
non-linearity in the detector response, manifested as $\pm5$~keV
shifts in the position of the known background peaks in the physics
data. We correct for these shifts by applying a 2nd-order polynomial
correction to the spectra of the reconstructed energies, binned in
1~keV intervals~\cite{kinr-1990-35_VIT}.

\begin{figure}[htbp]
\centering
\includegraphics[width=0.48\textwidth]{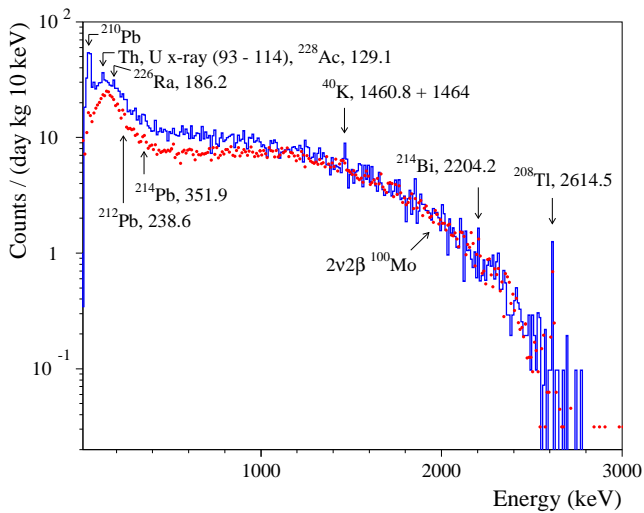}
\caption{Energy spectra accumulated with the \crystal~scintillating bolometers in the setup 1 (solid blue histogram, exposure $10.308$\,kg$\times$d) and setup 2 (red dotted histogram, exposure $31.927$\,kg$\times$d). The $\gamma$-ray energies are listed in keV.}
\label{fig:BG}
\end{figure}

The energy spectra of events acquired in setup 1 and setup 2 are shown in
Fig.~\ref{fig:BG}. Coincidences between the crystals exist but are
neglected in the analysis, taking into account a rather small
coincidence probability due to the detector positions in the setup
and a thick copper shield (minimum 2 mm) surrounding each
detector. A pulse-shape discrimination cut is applied to the
signals to find physical events and to reject pileup events; this
reduces tails in the resolution function. In
addition, $\alpha$ decays are eliminated from the spectrum using the
light-assisted particle identification cut, which achieves
about 9$\sigma$ $\alpha$/$\gamma$ separation
\cite{Poda:2017,Armengaud:2017b}. The light-assisted particle
identification removes not only fully contained $\alpha$ events from U/Th
chains, expected above 4 MeV, but also $\alpha$ decays with degraded
energies originating near the crystal surfaces. The
rate of such events is estimated to be
$0.1-0.2$ counts/yr/kg/keV in the $2.7-3.9$~MeV region.

The selection efficiency is found to be constant above $500$~keV,
and is evaluated to be
$(96.1\pm1.2)\%$ and $(96.6\pm0.7)\%$ for setups 1 and 2,
respectively. The exposure-weighted efficiency  for the complete data
set is $(96.5\pm0.6)\%$.
The selection efficiency estimate was
cross-checked using a prominent, but still low intensity, $\gamma$
peak of $^{40}$K resulting to a good agreement:
$(94.7\pm1.6)\%$.
Below 500 keV, the raw spectrum of triggered events
before the light yield and pulse shape selection has a significant
contribution from fake instrumental events.
We use events identified as $^{210}$Pb decays (a $46.5$~keV x-ray and
the corresponding $\beta$) to measure the selection efficiency of
$(90\pm10)\%$ at low energies.

\nopagebreak

\section{Results and discussion}

\subsection{Background model}
\label{sec:BG-model}

The most striking feature in the summed energy distribution in
Fig.~\ref{fig:BG} is the continuous spectrum characteristic of
\twonu\ decays, which dominates the data above $\sim1$~MeV.
The most prominent peaks in the spectra can be ascribed to contamination from
$^{40}$K and the daughters of the $^{238}$U and $^{232}$Th decay
chains.
The observed line shape of the
$^{40}$K peak is consistent with a subset of the events
containing the coincidence of the primary EC decay and subsequent
relaxation energy of the atomic shell with the $\gamma$-ray
event. This indicates two $^{40}$K sources: an external source far
from the detectors, and a source internal to the crystals. The ratios
of the other peaks to the
continuum indicates that the $\gamma$-line activity is dominated by
the external, far sources that are partially attenuated by the lead
and radiation shields of the cryostat. This conclusion is consistent
with the limits on the internal crystal contamination in $^{238}$U and
$^{232}$Th obtained from the analysis of the $\alpha$ region of the
energy spectrum.

Based on these observations, we construct a comprehensive background
model which includes a combination of ``internal'' (inside the
\crystal\ crystals), ``external'' sources (e.g. detector support
structures and the cryogenic vessels), and ``nearby'' sources
(surfaces close to the crystals, where one may expect a contribution
from $\beta$ events). The backgrounds are simulated using
the {\sc Geant4} package version 10.p03 (Livermore physics list)
\cite{Agostinelli:2003,Allison:2006,Allison:2016} with initial
kinematics given by the DECAY0 event generator
\cite{DECAY0a,DECAY0b}. The following ``external'' sources are simulated
on the 300 K cryostat vessel indicated in
Fig.~\ref{fig:Exp}:
\begin{itemize}
    \item $^{40}$K;
    \item $^{228}$Ac;
    \item late $^{232}$Th chain: $^{212}$Pb, $^{212}$Bi and
      $^{208}$Tl, assumed to be in secular equilibrium;
    \item late $^{238}$U chain: $^{214}$Pb and $^{214}$Bi in secular
      equilibrium;
    \item $^{137}$Cs, which was observed previously in the EDELWEISS
    setup~\cite{Armengaud:2017a,Scorza:2015vla}.
\end{itemize}
The following ``nearby'' sources are simulated in the materials near the
detectors:
\begin{itemize}
    \item $^{210}$Pb/$^{210}$Bi, assumed to be in secular equilibrium;
    \item $^{208}$Tl in the Kapton-based readout connectors, which are
      known to have measurable levels of contamination~\cite{Armengaud:2017a}.
\end{itemize}

The following ``internal'' sources are simulated inside the crystals:
\begin{itemize}
    \item $^{40}$K;
    \item $^{87}$Rb;
    \item $^{90}$Sr and $^{90}$Y;
    \item $^{210}$Pb/$^{210}$Bi;
    \item \twonu~decay of \isomo~to the ground state of \isoru;
    \item \twonu~decay of \isomo~to the first excited state of \isoru,
      $0^+$ at $1130.3$ keV. The half-life of this decay is fixed to the
      value determined by the NEMO-3
      collaboration~\cite{NEMO_excited}.
\end{itemize}
The $^{210}$Pb/$^{210}$Bi contribution is determined by the analysis of the
$^{210}$Po peaks in the $\alpha$-decay region of the energy spectrum.
The majority of the $^{210}$Pb/$^{210}$Bi/$^{210}$Po contamination is
attributed to the bulk of the crystals; this is also supported by the
shape of the $^{210}$Pb x-ray and $\beta$ spectra in the vicinity of
$46.5$~keV. A small contribution from the
``nearby'' sources (which appears primarily in setup 1) is treated as a
systematic uncertainty.

``Internal'' contamination of $^{40}$K and $^{87}$Rb in the bulk of the
crystals are added taking into account the observation of
$^{40}$K in some lithium molybdate crystals
\cite{Armengaud:2017b}, and similarity of lithium, potassium and
rubidium chemical properties. The presence of $^{90}$Sr-$^{90}$Y
in the crystals cannot be excluded; it is seen with marginal
significance in another bolometric
experiment~\cite{Azzolini:2019}.

A possibility of the full background reconstruction in a low
background experiment is limited by imprecise knowledge of the
locations of radioactive contaminations.
We build two models with different assumptions about the localization
of the background sources. In the default model, we simulate the full
geometry of the EDELWEISS cryostat including its payload, and assign
all of the ``external'' contamination to the 300~K vessel. As a
systematic check, we also develop a simplified model in which the
radioactive backgrounds are placed in copper shields of different
thickness around the crystal. This model is tuned to reproduce the
energy dependence of the observed intensities of the $\gamma$-peaks.

It should be stressed that no $\alpha$
decays from the U/Th chain, but few tens--hundreds $\mu$Bq/kg of
$^{210}$Po, were observed in the \crystal\ crystal
scintillators \cite{Poda:2017,Armengaud:2017b}, resulting in the
very stringent upper limits given in Table~\ref{tab:radcont}.
Therefore, bulk U/Th radioactivity of the crystals (except for the
contribution of $^{210}$Bi) is ignored in the background model,
  taking into account that the activity of
$^{100}$Mo in the crystals is at least three orders of magnitude
higher than the possible activity of U/Th daughters.

\begin{table}[htb]
\caption{Radioactive contamination of \crystal\
  crystal scintillators.
The limits are quoted at 90\% C.L.
}
\label{tab:radcont}
\begin{center}
\begin{tabular}{l|l|l|l}
 \hline
 Chain          & Radionuclide          & Activity      & Reference  \\
 ~              & ~                     &  (mBq/kg)     & ~ \\
 \hline
 ~              & $^{190}$Pt            & $\leq 0.003$  & \cite{Armengaud:2017b} \\
 $^{232}$Th     & $^{232}$Th            & $\leq 0.003$  & \cite{Armengaud:2017b} \\
                & $^{228}$Th            & $\leq 0.003$  & \cite{Poda:2017} \\
 $^{235}$U      & $^{235}$U             & $\leq 0.005$  & \cite{Armengaud:2017b} \\
 ~              & $^{231}$Pa            & $\leq 0.003$  & \cite{Armengaud:2017b} \\
 ~              & $^{227}$Ac            & $\leq 0.005$  & \cite{Armengaud:2017b} \\
 $^{238}$U      & $^{238}$U             & $\leq 0.005$  & \cite{Armengaud:2017b} \\
 ~              & $^{226}$Ra            & $\leq 0.003$  & \cite{Poda:2017} \\
 \hline
\end{tabular}
\end{center}
\end{table}
\normalsize

\begin{figure}[htbp]
\centering
 \includegraphics[width=0.48\textwidth]{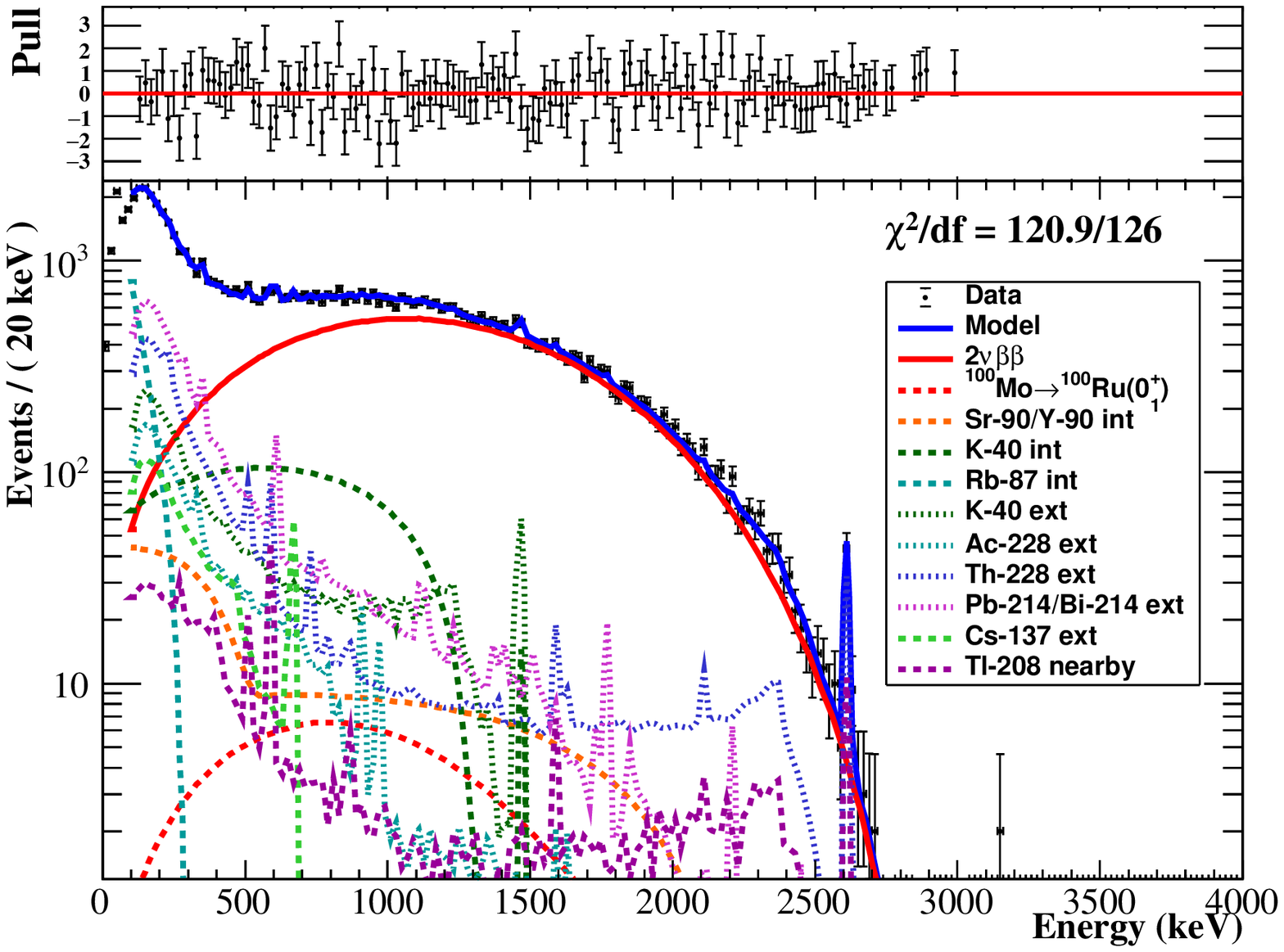}
\caption{Bottom: The energy spectrum accumulated with
\crystal\ scintillating bolometers (exposure is 42.235
kg$\times$d) and the fit in the energy range $[120-3000]$~keV. The
data points represent the data, the solid blue line shows the sum of
all components of the fit, the solid red line is the
\twonu\ contributions, and the other components of the fit are
described in the legend. Top: fit residuals normalized by the
statistical error of the data in each energy bin (``pulls'').
} \label{fig:2n2b}
\end{figure}

We constrain the background model and the \twonu\ half-life
by performing an extended maximum-likelihood fit~\cite{RooFit} to the
sum spectrum (the total exposure is 42.235 kg$\times$d, or
$3.798(9)\times10^{23}$ $^{100}$Mo nuclei$\times$yr), binned
uniformly with 20~keV bins. We perform a complementary
binned least-squares/maximum likelihood fit using PAW/MINUIT
software~\cite{PAW,MINUIT}; the two software packages return
consistent results.
The background model describes the data very well over a broad energy
range $[120-3000]$~keV (Fig.~\ref{fig:2n2b}). In order to assess the
sensitivity of the
background model and the \twonu\ half-life to the underlying
assumptions about the background composition, we vary the energy range
of the fit in 20 keV steps from 120 to 2000 keV (starting point) to
2300--3000 keV (final point), and find the value of the half-life
stable within the expected statistical variations.
The model, assuming the single-state
dominance mechanism of
the \twonu\ decay, describes the experimental data in the
$[120-3000]$~keV range with
$\chi^2=121$ for 126 degrees of freedom.

\nopagebreak

\subsection{Model of the \twonu\ decay}

We simulate
\twonu\ distributions using two
assumptions about the decay mechanism: the closure approximation
(in other words, high-state dominance, HSD), and the single-state
dominance (SSD) hypothesis. The SSD mechanism of \twonu\
decay was proposed in \cite{Abad:1984a} for nuclei where the $1^+$
ground state of intermediate nucleus may dominate the \twonu\
decay. $^{100}$Mo is one of a few cases where the SSD mechanism is
expected to have some merit
\cite{Griffiths:1992,Civitarese:1998,Simkovic:2001,Domin:2005,Kotila:2012}.
The data of the NEMO-3 experiment favor the SSD mechanism in
$^{100}$Mo \cite{Arnold:2019,Arnold:2004,Shitov:2006} and are
inconsistent with the HSD hypothesis.

\begin{figure}[htbp]
\centering
  \includegraphics[width=0.48\textwidth]{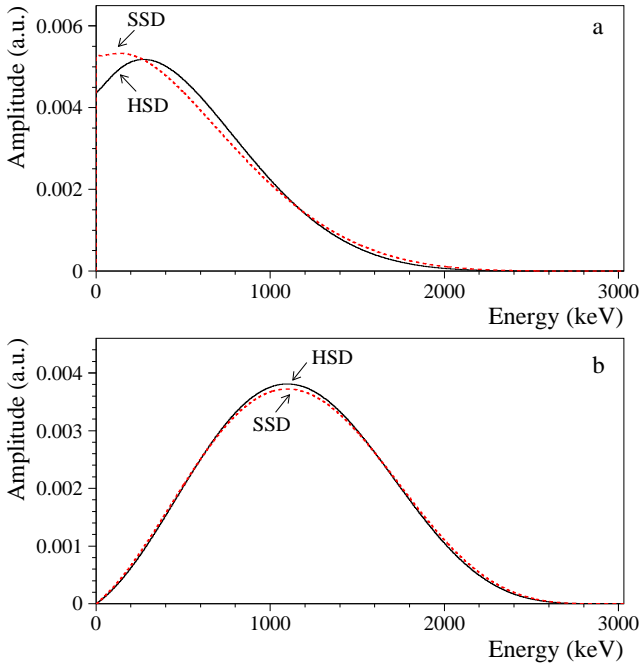}
\caption{Single-electron spectra (a) and summed energy spectra of two electrons (b) for the $^{100}$Mo$\rightarrow^{100}$Ru \twonu\ decay calculated in the HSD and SSD models. The spectra are normalized to unit area.}
 \label{fig:hsd-ssd}
\end{figure}

The energy
spectra of single electrons and summed two-electron energy spectra
for the $^{100}$Mo$\rightarrow^{100}$Ru \twonu\ decay using
calculations with the SSD and the HSD approximations
\cite{Kotila:2012} are shown in Fig. \ref{fig:hsd-ssd}. There is a
meaningful difference in the single-electron spectra for the HSD
and SSD models at low energies, while in the summed energy
spectra, measured by bolometric detectors, the difference is
substantially smaller. NEMO-3 analysis of the single-electron spectra
in $^{100}$Mo rules out the HSD hypothesis with high significance.

We use the SSD model of the \twonu\ decays in the baseline fit to the
experimental data, treating the difference between HSD and SSD models
as a systematic uncertainty (see Section~\ref{sec:syst}).

\nopagebreak

 The high statistics of the dataset, excellent resolution, and a high
signal-to-background ratio for energies above 1~MeV allow us to
test the spectral shape of the \twonu\ decays. We perform the fit in
the interval $[120-3000]$~keV range using the spectra generated under
the SSD and HSD hypotheses. The quality of both fits is
acceptable, but the HSD hypothesis returns a larger overall $\chi^2$
by $12.5$ units (the negative log-likelihood is larger for the HSD
hypothesis by $8.2$ units) .

Since $\sqrt{\Delta\chi^2}$ is not, strictly speaking, equal to the
significance of discriminating one hypothesis over
another~\cite{ref:Chi2Statistic}, we use an ensemble of 10,000
pseudo-experiments to determine the confidence level at which the SSD
hypothesis is preferred over the HSD hypothesis. In each
pseudo-experiment, we generate the energy distribution of signal and
background events from the probability density functions returned by
the fit to the $[120-3000]$~keV range. The events are generated using
the SSD hypothesis, and then two fits using the SSD and HSD hypotheses
are performed. From this ensemble, we determine the mean of the
expected distribution of the log-likelihood ratio
$\log(\mathcal{L_\mathrm{HSD}}/\mathcal{L_\mathrm{SSD}})$
($\mu=+8.04$), its standard deviation ($\sigma=2.68$), and the
probability for the ratio
$\log(\mathcal{L_\mathrm{HSD}}/\mathcal{L_\mathrm{SSD}})$ to fluctuate
above zero ($p=0.0014\pm0.0004$). Similar values are obtained for an
ensemble of pseudo-experiments randomly sampled from the energy
spectrum observed in the data.
We interpret these results as a preference for the SSD
hypothesis over HSD with the statistical significance of $>3\sigma$.

\subsection{Half-life of $^{100}$Mo}

The background model described above is sensitive to the exact
composition and location of the background sources. Since several possible
background sources have broad energy spectra similar to \twonu, the
correlations between the background source activities and the
\twonu\ half-life are significant. When fit over the broad
energy range, e.g. $[120-3000]$~keV, the best-fit \twonu\
half-life value has a small statistical uncertainty, but a large
systematic uncertainty due to the model of background composition and
location, as well as the reconstruction efficiency uncertainty at low energies.

For these reasons, we determine the $^{100}$Mo half-life by fit the spectrum
in the reduced energy range $[1500-3000]$~keV. In this range, only two
background contributions are relevant: the late-chain $^{232}$Th
decays from external sources, dominated by the $2615$~keV $^{208}$Tl
$\gamma$ line
and its Compton continuum and the
late-chain $^{238}$U decays from external sources, dominated by
$^{214}$Bi and its Compton
continuum. For completeness, we include a possible contribution from
$^{228}$Ac $\gamma$ spectra from external sources, and a possible
contribution from internal $^{90}$Sr-$^{90}$Y $\beta$-decays. The
max-likelihood values of both of those
components are consistent with zero. We also split the $^{208}$Tl
component into ``external'' and ``nearby'' sources.
All background components of the fit are restricted to the physical
(positive yield) range.

The interval $[1500-3000]$~keV contains $23.5\%$ of the
\twonu\ spectrum. 9183 events are found in this range in the
$42.235$\,kg$\times$d of exposure, with 91\% attributed to
\twonu\ events.
The fit quality is excellent ($\chi^2=50$ for 61 degrees of freedom)
with modest (80\%) correlations between the \twonu\ half-life and the
background components. The fit returns
$8370^{+162}_{-214}\,(\mathrm{stat.})$ \twonu\
events in the fit region (extrapolated to the full energy range,
the number of \twonu\ events is $35638^{+693}_{-912}\,(\mathrm{stat.})$).
Taking into account the selection efficiency ($0.9646\pm0.0060$), we
find the half-life
$\THL=[7.12^{+0.18}_{-0.14}\,(\mathrm{stat.})] \times 10^{18}$ yr.
The uncertainties are asymmetric due to the correlations
with the background components that are consistent with zero and are
restricted to the physical (positive) yield, most notably $^{90}$Y.

For comparison, the energy interval $[120-3000]$~keV contains $63717$
events; the fit attributes $35405\pm605$ to \twonu\ ($99.4\%$ of the
\twonu\ spectrum is contained in the $[120-3000]$~keV interval). We
find
$\THL=[7.13\pm0.12\,(\mathrm{stat.})\pm0.20\,(\mathrm{syst.})]\times10^{18}$~yr
for this interval, in excellent agreement to the fit to the more
restricted range.
The wide energy interval is susceptible to larger
systematic uncertainties (discussed below), so we consider this fit as
a cross-check.

\subsection{Systematic uncertainties}
\label{sec:syst}

We vary the underlying assumptions in the default fit over the energy
range $[1500-3000]$~keV to evaluate the systematic
uncertainties.
Signal efficiency contributes
$0.6\%$ to the systematic error on \THL. Uncertainty in the energy
scale contributes $0.2\%$.
Variation of the bin width (from 10~keV to 30~keV)
change \THL\ by up to $0.8\%$. We attribute this variation to the
uncertainty in the resolution function applied to the simulated
background spectra, and treat the difference as the systematic
error.

As it was already mentioned, the internal contamination of the
\crystal\ crystal scintillators by U/Th is very low.
Assuming the activities of the $\beta$ active daughters of
$^{232}$Th ($^{228}$Ac, $^{212}$Pb, $^{212}$Bi, $^{208}$Tl) and
$^{238}$U ($^{234m}$Pa, $^{214}$Pb, $^{214}$Bi, $^{210}$Bi) to be
equal to the activity limits (see Table~\ref{tab:radcont}), the
total contribution of the bulk radioactivity is $\leq 0.1\%$ in the
region of the fit. The contribution of cosmic
muons was estimated on the basis of the measurements with
germanium bolometers by the EDELWEISS collaboration
\cite{Schmidt:2013,Kefelian:2016} and the simulations of the muon
induced background in germanium detectors taking into account the
muon flux as a function of slant depth \cite{Mei:2006}. A contribution
of cosmic-muons background is estimated to be less than 14 counts
($\leq 0.15\%$). We treat these backgrounds as systematic uncertainty
($0.2\%$).
In order to further test the sensitivity to the assumptions about the
background composition, we repeat the fit after removing the
background components consistent with zero activity ($^{90}$Sr and
$^{228}$Ac). As expected, the value of \THL\ determined in the
$[1500-3000]$~keV interval changes very little ($0.1\%$).

We study the effects of the localization of the sources by
comparing fits with two independent sets of simulated spectra: one
using the complete EDELWEISS geometry and placing all ``external''
sources on the 300~K vessel, and a simplified detector geometry with
location of the sources tuned to reproduce the
energy dependence of the observed intensities of the
$\gamma$-peaks (0.8\%).
We test the sensitivity to the temporal and spatial
variations in the background conditions by splitting the dataset into
five independent subsets of similar exposure: setup 1, and 4 separate
crystals in setup 2. The five datasets agree within the statistical
uncertainties with the half-life determined from the summed spectrum
($\chi^2=2.6$ for 4 degrees of freedom). We conclude that there is no
evidence for an additional systematic uncertainty arising from this
test~\cite{Barlow:2002}.

The HSD decay model changes \THL\ by $0.4\%$; we
consider this difference to be a conservative upper limit on the systematic
error induced by the uncertainty in \twonu\ spectral shape.
The description of the \twonu\ energy
spectrum can be refined using the improved formalism of the
two-neutrino double-beta decay calculations \cite{Simkovic:2018}.
We should
note that like all other measurements of \twonu\ half-life, our
\twonu\ decay model does not currently include
$\mathcal{O}(\alpha)$ and $\mathcal{O}(\alpha Z)$ radiative
corrections other than the Coulomb (final state) corrections computed
in Ref.~\cite{Kotila:2012}.

Finally, we account for uncertainties in the number of $^{100}$Mo
nuclei, the live-time of the measurement, finite Monte Carlo
statistics, and the rate of the
\twonu\ decay to the first $0^+$ excited level of $^{100}$Ru.
The summary of the
systematic uncertainties is given in Table~\ref{tab:syst}.

\begin{table}[htb]
\caption{Estimated systematic uncertainties (\%).}
\begin{center}
\begin{tabular}{l|l}
 \hline
Binning of the energy spectrum          & $0.8$ \\
  \hline
 Localization of radioactive sources        & $0.8$ \\
 \hline
 Selection efficiency               & $0.6$ \\
 \hline
\twonu\ spectral shape              & $0.4$ \\
 \hline
 Monte Carlo statistics             & $0.4$ \\
 \hline
 Background composition                         & $0.2$  \\
 \hline
 Exposure of $^{100}$Mo                           & $0.2$ \\
 \hline
 Energy scale                                   & $0.2$  \\
 \hline
 $\THL(^{100}\mathrm{Mo}\to^{100}\mathrm{Ru}(0^+_1))$ & $0.1$  \\
 \hline

 Total systematic error             &  $1.4$ \\
 \hline
\end{tabular}
\end{center}
\label{tab:syst}
\end{table}

\section{Summary}

\begin{figure}[htbp]
\centering
  \includegraphics[width=0.48\textwidth]{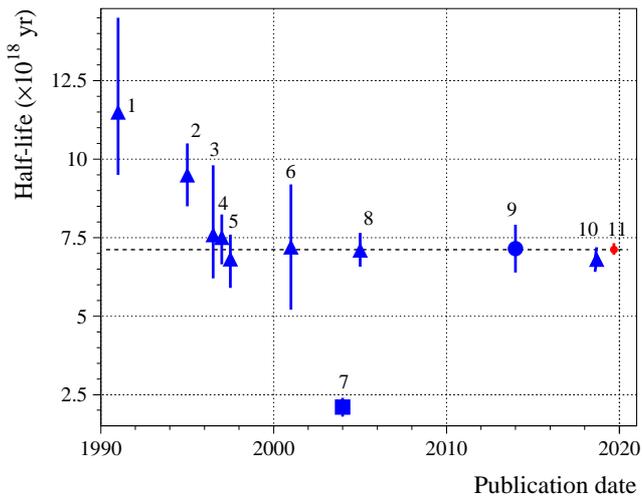}
\caption{A historical perspective of \THL\ of
$^{100}$Mo as a function of the publication date in the
experiments: (1) ELEGANT V \cite{Ejiri:1991}, (2) NEMO-2
\cite{Dassie:1995}, (3) segmented Si(Li) detector
\cite{Alston:1997}, (4) NEMO-2 reanalyzed \cite{Vareille:1997},
(5) Hoover Dam using a time-projection chamber
\cite{DeSilva:1997}, (6) DBA (liquid argon detector)
\cite{Ashitkov:2001}, (7) geochemical experiment
\cite{Hidaka:2004}, (8) preliminary result of NEMO-3
\cite{Arnold:2005}, (9) low temperature ZnMoO$_4$ bolometers
\cite{Cardani:2014}, (10) final result of NEMO-3
\cite{Arnold:2019}, (11) present study.}
 \label{fig:hist}
\end{figure}

Adding all systematic contributions in quadrature, the
half-life of $^{100}$Mo relative to the \twonu\ decay to the
ground state of $^{100}$Ru is:
$$
\THL=[7.12^{+0.18}_{-0.14}\,(\mathrm{stat.})\pm0.10\,(\mathrm{syst.})]\times10^{18}~\mathrm{yr}.
$$
This that can be simplified further by summing in quadrature the
systematic and statistical errors:
$$
\THL = (7.12^{+0.21}_{-0.17})\times 10^{18}~\mathrm{yr}.
$$
The half-life value is in an agreement with all the counting
experiments after 1995 (a history of $^{100}$Mo half-lives is
shown in Fig.~\ref{fig:hist}).

\nopagebreak

The precision of the present result is higher thanks to
the certain advantages of the CUPID-Mo detection technique based on lithium molybdate
scintillating bolometers produced from isotopically enriched
$^{100}$Mo. The measurement features a high and accurately defined
detection efficiency (particularly, because there is no fiducial volume
uncertainty), a high energy resolution that allows building an
accurate background model, a very low radioactive contamination
of the crystal scintillators and of the EDELWEISS-III cryostat.
The very low-background conditions, together with utilization of
enriched $^{100}$Mo, allowed us to reach a rather high signal to
background ratio (approximately 10:1).

An effective nuclear matrix element for \twonu\ decay of
$^{100}$Mo to the ground state of $^{100}$Ru, assuming the SSD
mechanism, can be calculated as $|M^\mathrm{eff}_{2\nu}|=0.184^{+0.002}_{-0.003}$
by using the phase-space factor $4134\times10^{-21}$ yr$^{-1}$
calculated in \cite{Kotila:2012}. The effective nuclear matrix
element can be written as product $M^\mathrm{eff}_{2\nu}=g_{A}^2\times
M_{2\nu}$, where $g_{A}$ is axial vector coupling constant,
$M_{2\nu}$ is nuclear matrix element. While the value of
$M_{2\nu}$ is almost independent on the $g_A$ and can be
calculated with a reasonable accuracy, the possible range of $g_A$
can be quenched from $1.2694$ (the free nucleon value) to 0.6--0.8
\cite{Barea:2013,Robertson:2013,DellOro:2014,Suhonen:2017}.

Taking into  account that $^{100}$Mo nuclei decay by the two
modes: to the ground state and to the first $0^+$ excited level of
$^{100}$Ru, the actual half-life of $^{100}$Mo (using the most
accurate measurement of the decay of $^{100}$Mo to the first $0^+$
1130.3 keV excited level of $^{100}$Ru~\cite{NEMO_excited} is:
\begin{center}
$T_{1/2}=(7.05^{+0.21}_{-0.17})\times 10^{18}$~yr.
\end{center}
In other words, the branching ratios are $99.06(11)\%$
and $0.94(11)\%$ for the \twonu\ decay of $^{100}$Mo to the
ground state and to the first $0^+$ $1130.3$ keV excited level of
$^{100}$Ru, respectively.

\section{Conclusions}

The two-neutrino double-beta decay of $^{100}$Mo to the ground
state of $^{100}$Ru is measured precisely with four
$^{100}$Mo-enriched highly radiopure lithium molybdate
scintillating bolometers $\approx0.2$ kg each operated in the
EDEL\-WEISS-III low background setup at the Modane underground
laboratory (France). The $^{100}$Mo half-life value
$\THL=7.12^{+0.18}_{-0.14}\,(\mathrm{stat.})\pm0.10\,(\mathrm{syst.})]\times10^{18}$~yr
is measured with 42.235 kg$\times$d
exposure. The measurement, performed in the energy range
$[1500-3000]$~keV is statistics-limited, and can be further improved
with more data.
The result, being in a
good agreement with all previous counting experiments after 1995,
is the most accurate determination of the $^{100}$Mo half-life.

Moreover, the half-life value measured with the relative uncertainty
of $^{+2.9}_{-2.4}\%$ is among the most precise
measurements of any \twonu\ decay to date.
Other leading measurements are
of $^{130}$Te by CUORE (2.8\%
\cite{Caminata:2019}), $^{136}$Xe by EXO-200 (2.8\%
\cite{Albert:2014}) and KamLAND-Zen (3.3\%
\cite{Gando:2016,Gando:2019}), $^{76}$Ge by GERDA (4.9\%
\cite{Agostini:2015}), $^{116}$Cd by Aurora (5.3\%
\cite{Barabash:2018}), $^{82}$Se by NEMO-3 (6.4\%
\cite{Arnold:2018}) and
CUPID-0 ($^{+2.2}_{-1.6}\%$ \cite{Azzolini:2019}),  $^{150}$Nd by NEMO-3 (7.1\%
\cite{Arnold:2016}), $^{96}$Zr by NEMO-3 (8.9\%
\cite{Argyriades:2010}) and other observations of \twonu\
decay in $^{48}$Ca, $^{128}$Te, and $^{238}$U ($\approx$10--30\%,
e.g. see in \cite{Barabash:2019}).
The three of four most precise measurements of the \twonu\ half-life are from
the bolometric experiments, demonstrating the power of the technique.

The high precision of the measurement is achieved thanks to
utilization of enriched detectors with an extremely low level of
radioactive contamination, operated in the low background
environment deep underground. A rather high signal to background ratio in the energy interval of the analysis is reached.
The calorimetric approach, together with an excellent energy
resolution of the \crystal\ detectors, ensured a high,
clearly defined detection efficiency, and accurate background
reconstruction, that are typically the main sour\-ces of
systematic error in the \twonu\ measurements.

In agreement with the observation by
NEMO-3~\cite{Arnold:2019,Arnold:2004,Shitov:2006}, we favor the SSD
mechanism of the \twonu\ decay over the HSD mechanism, with the
statistical significance of $>3\sigma$.
Therefore, we derive the half-life assuming the SSD mechanism of
the decay. An effect of the energy spectra shape due to the
different mechanisms of the decay is included in the systematic
error of the half-life.

The half-life and the spectral shape accuracy are expected to be
further improved in the CUPID-Mo experiment~\cite{Armengaud:2019}
running now in its
first phase with 20 enriched \crystal\ scintillating
bolometers (with mass $\approx 0.2$ kg each).

\begin{acknowledgements}

The authors gratefully thank Prof. F. Iachello for communications
related to the single state dominance in $^{100}$Mo.  This work has been
partially performed in the framework of the LUMINEU program, a project
funded by the Agence Nationale de la Recherche (ANR, France). The help
of the technical staff of the Laboratoire Souterrain de Modane and of
the other participant laboratories is gratefully acknowledged.
The group from Institute for Nuclear Research
(Kyiv, Ukraine) was supported in part by the IDEATE International
Associated Laboratory (LIA). A.S. Barabash, S.I. Konovalov,
I.M. Makarov, V.N. Shlegel and V.I. Umatov were supported by Russian
Science Foundation (grant No.  18-12-00003). O.G. Polischuk was
supported in part by the project ``Investigations of rare nuclear
processes'' of the program of the National Academy of Sciences of
Ukraine ``Laboratory of young scientists''. The Ph.D. fellowship of
H. Khalife has been partially funded by the P2IO LabEx
(ANR-10-LABX-0038) managed by the ANR (France) in the framework of the
2017 P2IO Doctoral call.
J.~Kotila is supported by Academy of Finland (Grant No. 314733).
C. Rusconi is supported by the National
Science Foundation Grant NSF-PHY-1614611. This material is also based
upon work supported by the US Department of Energy (DOE) Office of
Science under Contract No. DE-AC02-05CH11231; by the DOE Office of
Science, Office of Nuclear Physics under Contract
Nos. DE-FG02-08ER41551 and DE-SC0011091; by the France-Berkeley Fund,
the MISTI-France fund, and by the Chateaubriand Fellowship of the
Office for Science \& Technology of the Embassy of France in the United
States.

\end{acknowledgements}

\end{document}